\documentclass[amsmath,amssymb,aps,prl,twocolumn,superscriptaddress]{revtex4-2}
\usepackage{graphicx}
\usepackage{bm}

\begin{document}


\title{Elastic chiral Landau level and snake states in origami metamaterials}


\author{Shuaifeng Li}
\affiliation{Department of Aeronautics and Astronautics, University of Washington, Seattle, WA 98105, USA}
\author{Panayotis G.Kevrekidis}
\affiliation{Department of Mathematics and Statistics, University of Massachusetts, Amherst, MA 01003, USA}
\author{Jinkyu Yang}
\affiliation{Department of Aeronautics and Astronautics, University of Washington, Seattle, WA 98105, USA}
\affiliation{Department of Mechanical Engineering, Seoul National University, Seoul 08826, Republic of Korea}


\begin{abstract}
In this study, we present a method for generating a synthetic gauge field in origami metamaterials with continuously varying geometrical parameters. By modulating the mass term in the Dirac equation linearly, we create a synthetic gauge field in the vertical direction, which allows for the quantization of Landau levels through the generated pseudomagnetic field. Furthermore, we demonstrate the existence and robustness of the chiral zeroth Landau level. The unique elastic snake state is realized using the coupling between the zeroth and the first Landau levels. Our results, supported by theory and simulations, establish a feasible framework for generating pseudomagnetic fields in origami metamaterials
with potential applications in waveguides and cloaking.
\end{abstract}


\maketitle

\textit{Introduction.}--The physical realization of the synthetic gauge field $\bm{A}$ has been an engaging field of condensed matter physics research as it provides an additional degree of freedom for modulating waves~\cite{cooper2019topological,fang2013controlling,lin2014light,yang2021demonstration,lu2014topological,khanikaev2017two,yang2015topological}. The resulting pseudomagnetic fields $\bm{B}=\nabla\times\bm{A}$ can induce a nontrivial band topology. Specifically, in phononic systems, the inertness of the elastic waves to the genuine magnetic field makes the realization of pseudomagnetic fields especially demanding for inducing intriguing phenomena such as Landau quantization and artificial Lorentz forces. 

In previous studies, the deformation of the artificial elastic periodic structures has been used to shift the Dirac cone in the $k_{x}$-$k_{y}$ momentum space, creating the synthetic gauge field in the $x$-$y$ plane and a corresponding pseudomagnetic field perpendicular to the $x$-$y$ plane~\cite{abbaszadeh2017sonic,peri2019axial,yang2017strain}. In addition, the gradient variation of geometrical parameters of structures can modulate the position of the Dirac cone in the $k_{x}$-$k_{y}$ momentum space, which offers a simple and practical way to realize the pseudomagnetic fields~\cite{brendel2017pseudomagnetic,wen2019acoustic,yan2021pseudomagnetic}. The resulting pseudomagnetic field facilitates the observation of various magnetic-field-related phenomena in phononic systems such as Landau plateaus and quantum-Hall-like edge states~\cite{abbaszadeh2017sonic,yang2017strain,yan2021pseudomagnetic,brendel2017pseudomagnetic,wen2019acoustic}. Nevertheless, there is a lack of the exploration of out-of-plane synthetic gauge field~(in-plane pseudomagnetic field), which may result in the distinctive form of Landau quantization. Furthermore, an intriguing transport phenomenon, namely a snake state, has been realized in the on-chip structures with the opposite pseudomagnetic field, but achieving opposite pseudomagnetic field needs the sophisticated design and fabrication~\cite{yan2021pseudomagnetic}. A better strategy is still in its infancy and worth exploration.

Origami, an ancient art of paper folding, has recently attracted the 
attention of the physics and engineering communities due to its remarkable potential for various applications. Origami metamaterials, which are composed of origami elements with intricate folding techniques, offer unique kinematic motions, functionalities, and mechanical properties that are not found in traditional materials~\cite{lv2014origami,filipov2015origami,wickeler2020novel}. These properties make origami metamaterials ideal for use in diverse applications, such as aerospace ones~\cite{zirbel2013accommodating}, architected materials~\cite{del2010adaptive,lyu2021origami}, and biomedical devices~\cite{miyashita2015untethered,edmondson2013oriceps,kuribayashi2006self}. In addition to static and quasi-static mechanical properties, origami metamaterials also possess rich dynamics that can be utilized to develop prospective devices for impact mitigation and vibration control, revealing the potentials in revolutionizing these fields~\cite{zhou2016dynamic,yasuda2020data,yasuda2019origami,miyazawa2022topological}. Furthermore, the folding patterns and geometric properties of origami metamaterials have shown significant potential in condensed matter physics, particularly in the study of topological states~\cite{chen2016topological,miyazawa2022topological,li2023origami}. The unique mechanical properties of origami metamaterials, coupled with their ability to be easily adjusted through initial configurations and crease pattern engineering, make them the ideal platforms to produce synthetic gauge fields, and thereby generate pseudomagnetic fields, where the gradient of system properties is the key ingredient~(geometry gradient, modulus gradient, etc.). This opens up intriguing possibilities for exploration in origami metamaterials, which have yet to be investigated.

In our work, we introduce an origami metamaterial composed of Kresling origamis arranged in a honeycomb lattice bearing a gradient in geometrical parameters, as explained below. By appropriately designing the height of the origami, we generate a synthetic gauge field along the vertical direction through a linear variation of the Dirac effective mass, leading to the formation of a pseudomagnetic field. With the application of this pseudomagnetic field, we observe the quantization of Landau levels, where the robustness of the zeroth Landau level is confirmed by the weak backscattering of chiral elastic wave propagation against obstacles. Furthermore, we use the coupling between the zeroth and first Landau levels to demonstrate the elastic snake state, where the wavy propagation trajectory of elastic waves enables the bypassing of obstacles. Our work, supported by the excellent agreement of numerical calculations and theoretical analysis, provides a promising path towards achieving chiral Landau levels in origami metamaterials, with potential applications in origami-based architectures for the sophisticated manipulation of elastic waves.

\begin{figure}[h]
    \includegraphics[width=0.5\textwidth]{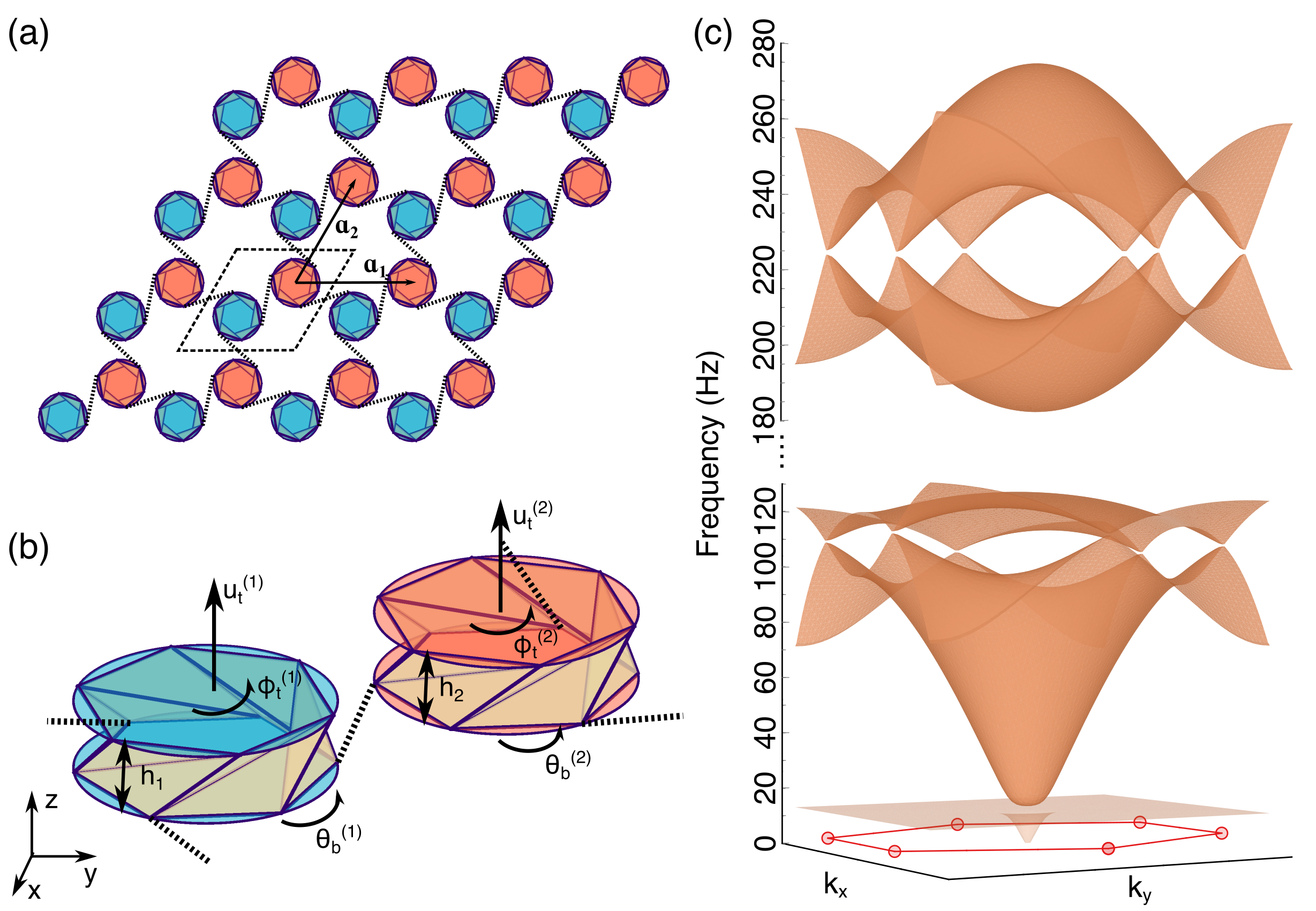}
    \caption{Design of the origami metamaterials.
    (a) The top view of the origami metamaterials with the Kresling origamis arranged in the honeycomb lattice. The origamis are connected by the reverse torsional spring. $\bm{a_{1}}$ and $\bm{a_{2}}$ denote two basis vectors of the unit cell enclosed by the black dashed line.
    (b) The side view of the unit cell composed of two origamis with initial heights $h_{1}$ and $h_{2}$ and initial rotaional angle $\theta_{0}=70^{\circ}$. Each origami has three degrees of freedom $u_{t}$, $\phi_{t}$ and $\theta_{b}$ with the bottom disk pinned to the ground.
    (c) The band structure calculated from the linearized truss model of origami. The Dirac cone emerges in six corners of the Brillouin zone, which is marked as red line.\label{fig:fig1}
    }
\end{figure}

\textit{The design of origami metamaterials.}--As shown by the top view in FIG.~\ref{fig:fig1}(a), we design our origami metamaterials by coupling Kresling origamis in the honeycomb lattice using reversed torsional springs $k_{c}$. The introduction of $k_{c}$ can induce opposite torque in the connected units~\cite{wu2018dial,pal2016helical}. The bottom plates are pinned to the ground. The primary unit cell is chosen to be a rhombus enclosed by the black dashed line with basic vector $\bm{a_{1}}=(\sqrt{3}a,0)$ and $\bm{a_{2}}=(\frac{\sqrt{3}}{2}a,\frac{3}{2}a)$, where $a=100~\mathrm{mm}$ is the side length of the honeycomb. FIG.~\ref{fig:fig1}(b) displays the side view of the unit cell. The design parameters and mechanical parameters are shown in Supplemental Material. Therein, the initial heights are denoted as $h_{1}$ and $h_{2}$, while the initial rotation angles are kept the same as $\theta_{0}=70^{\circ}$. Here, the initial rotation angle $\theta_{0}$ indicates the initial angle difference between bottom plate and top plate after fabrication. Each origami has three degrees of freedom: axial displacement of the top plate $u_{t}$, rotational displacements of the top plate $\phi_{t}$ and bottom plate $\theta_{b}$. Note that here rotational displacement of the bottom plate $\theta_{b}$ should not be confused with the initial rotation angle in the origami configuration $\theta_{0}$.

When the heights of two origamis are the same ($h_{1}=h_{2}=20~\mathrm{mm}$), the spatial inversion symmetry is preserved. As shown in FIG.~\ref{fig:fig1}(c), the associated band structure featuring six bands is calculated using the linearlized truss model of Kresling origami~(see Supplemental Material)~\cite{yasuda2017origami,li2023origami,miyazawa2022topological}. Using the discrete model of the origami structure, we pursue the effective wave motion in the finite wavenumber and finite frequency, and analyze origami behaviors near spectral singularities~(Dirac cone). The lowest two bands are almost flat except $\bm{k}=0$, within the frequency range of $12~\mathrm{Hz}$ to $14~\mathrm{Hz}$ and are illustrated in high transparency because they are not considered in our discussion below. The third and the fourth bands, the fifth and sixth bands degenerate to form the Dirac cones at lower frequency and higher frequency. Here, the problem of determining the dispersion relation in each frequency range can be cast in the form of a $2\times 2$ eigenvalue problem associated with the following Hamiltonian:
\begin{eqnarray}
    \label{equ:1}
    H=v(k_{x}\sigma_{x}\pm k_{y}\sigma_{y})
\end{eqnarray}
where $v$ is the Dirac velocity at $(k_{x}, k_{y})$, and $\sigma_{x}$ and $\sigma_{y}$ are Pauli matrices. This Hamiltonian maps to the massless Dirac equation associated with locally linear dispersion. When the heights of two origamis are not the same, the spatial inversion symmetry is broken since the coupling behaviors of Kresling origami between axial and rotational displacements are changing as a function of height~(Supplemental Material). This broken spatial inversion symmetry essentially introduces a $\sigma_{z}$ component into the Hamiltonian which can be expressed as:
\begin{eqnarray}
    \label{equ:2}
    H=v(k_{x}\sigma_{x}\pm k_{y}\sigma_{y})+m_{K}\sigma_{z}
\end{eqnarray}
The effective mass term $m_{K}$ will break the Dirac degeneracy and open the band gap locally. The theoretically predicted width of the band gap is $2|m_{K}|$. Therefore, we can expect that the Dirac cone at lower and higher frequency will open up when the heights for two origamis are not the same, which will be elaborated below to show its important role for the formation of the synthetic gauge field. A similar band gap opening mechanism is through the bistability of Stewart platform in the honeycomb lattice to realize topological phase transition~\cite{wu2018dial}.

\begin{figure}[h]
    \includegraphics[width=0.5\textwidth]{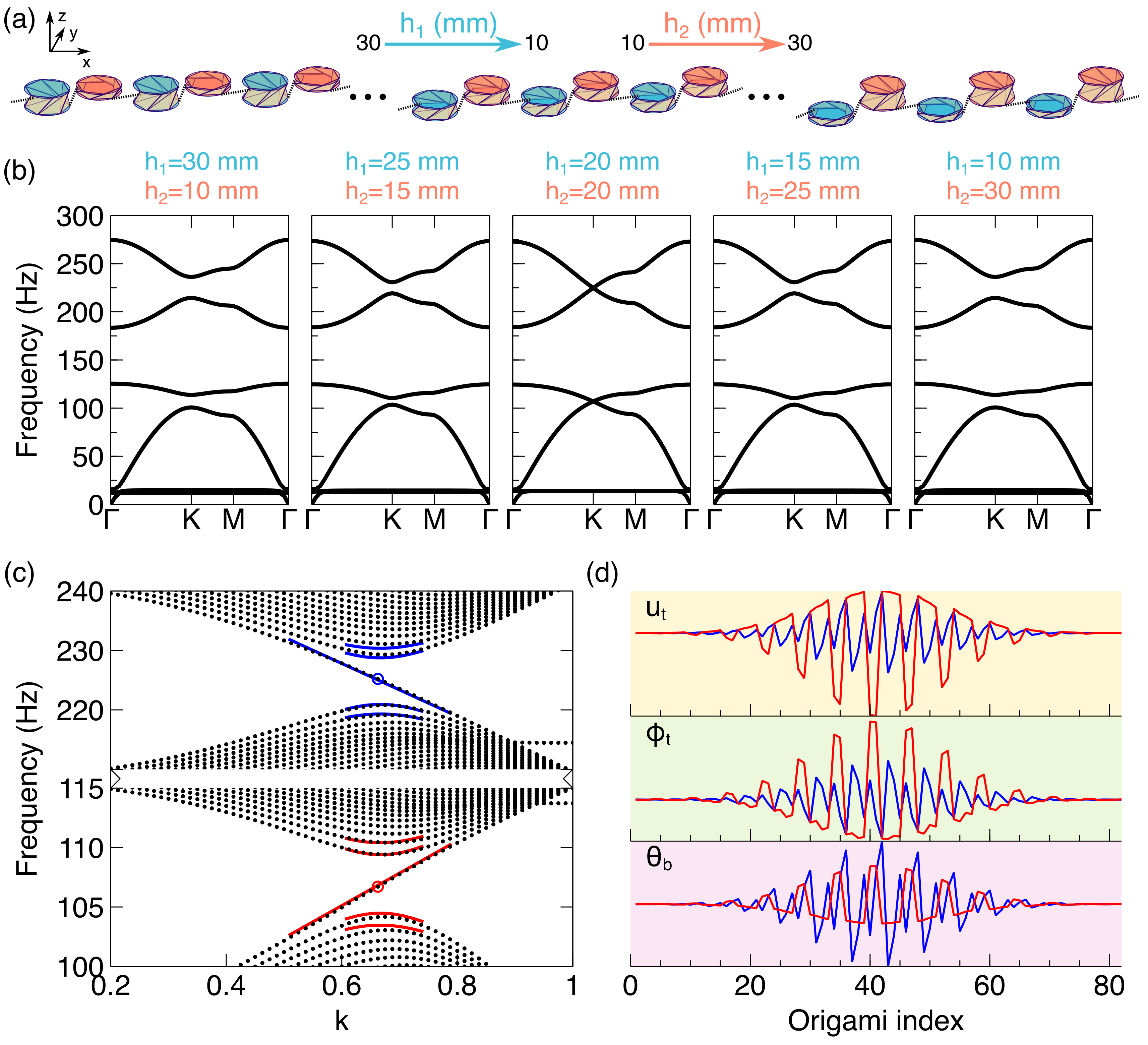}
    \caption{Synthetic gauge field and Landau level.
    (a) The schematic of the supercell composed of $41$ unit cells (82 origamis) with height gradient. The initial height of origami $h_{1}$, $h_{2}$ are linearly varying from $30~\mathrm{mm}$ to $10~\mathrm{mm}$ and from $10~\mathrm{mm}$ to $30~\mathrm{mm}$, respectively.
    (b) The band structure along $\Gamma$-$K$-$M$-$\Gamma$ calculated by the unit cell with different configurations.
    (c) The projected band structure of the supercell under the periodic boundary condition along $y$ direction. Two ends of the supercell are connected to the wall. The dotted lines denote the calculated Landau levels, while the solid lines denote the theoretical Landau levels.
    (d) The eigenmodes representing $u_{t}$, $\phi_{t}$ and $\theta_{b}$ are from top to bottom panel. Red and blue lines correspond to the red and blue circles in (c).
    \label{fig:fig2}}
\end{figure}

\textit{Emergence of elastic Landau levels.}--We study the evolution of the band structure when the height difference of two origamis $\Delta h=h_{1}-h_{2}$ varies from $20~\mathrm{mm}$ to $-20~\mathrm{mm}$, while keeping $h_{1}+h_{2}=40~\mathrm{mm}$. This variation and such configuration in the form of the supercell with $41$ unit cells is shown in FIG.~\ref{fig:fig2}(a). The band structures along $\Gamma-K-M-\Gamma$ for the corresponding configurations are displayed from left to right in FIG.~\ref{fig:fig2}(b). The band gaps between the third and the fourth bands, the fifth and sixth bands experience the open-close-reopen process. The formation of a band gap at the vicinity of $\bm{K}$ point is due to the mass term in the Dirac equation as explained in Equation~(\ref{equ:2}). Besides, the larger band gap results from the larger mass term, which can be theoretically calculated by the $\bm{k}\cdot \bm{p}$ perturbation method~(see Supplemental Material)~\cite{slonczewski1958band,chen2019topological,mousavi2015topologically,mei2012first,lu2014dirac}. The band gap and resulting mass term at the lower and higher frequency of the corresponding configuration in FIG.~\ref{fig:fig2}(a) are discussed in the Supplemental Material, and are featured by the linear variations of the width of band gap and mass term $m_{K}$ along the $x$ direction.

With this configuration, one can obtain an effective Hamiltonian:
\begin{eqnarray}
    \label{equ:3} H=v(\hat{k}_{x}\sigma_{x}\pm\hat{k}_{y}\sigma_{y})+m_{K}(x)\sigma_{z}
\end{eqnarray}
where $\hat{k}_{x}$ and $\hat{k}_{y}$ are the wave vector operators. Therein, $\hat{k}_{x}=-i\partial_{x}$ with the translational symmetry being broken
along the $x$ direction, while $\hat{k}_{y}=k_{y}$ with the translational
symmetry being preserved with periodic boundary condition along the $y$ direction. Besides, the mass term $m_{K}$ in the Dirac Hamiltonian is linear with respect to $x$~($m_{K}=qx$) and the signs of $q$ are opposite at lower and higher frequency~(Supplemental Material). According to the form of the Hamiltonian, a vector potential along the $z$ direction $A_{z}=m_{K}(x)$ is introduced, suggesting an effective canonical momentum $\hat{k}_{z}=k_{z}+A_{z}$~($k_{z}=0$) in the system. Hence, we can expect the in-plane pseudomagnetic field $B_{y}=\nabla\times A_{z}$. Note that previous studies focus on the synthetic gauge field in the $x$-$y$ plane, resulting in the out-of-plane pseudomagnetic field. For our origami metamaterial, the uniform pseudomagnetic field of magnitude $B_{y}=12.5~\mathrm{T}$ and $B_{y}=19.5~\mathrm{T}$ can be constructed for lower frequency and higher frequency, respectively, where the directions of pseudomagnetic field are opposite. With such a pseudomagnetic field affecting our system, the energy levels will be quantized as below~(see theoretical derivations in the Supplemental Material):
\begin{eqnarray}
    \label{equ:4}
    \omega_{n}=\begin{cases}
    \mathrm{sgn}(q)vk_{y},& n=0\\
    \pm\sqrt{v^2k^2_{y}+2n|q|v},& n\geq 1
    \end{cases}
\end{eqnarray}
where $\pm$ correspond to the $\bm{K}$ and $\bm{K}'$ valley, respectively. Here, we derive the dispersion relation with the chiral Landau levels in a two-dimensional system, whereas previous discussions on such chiral Landau levels are based on three-dimensional Weyl systems~\cite{pikulin2016chiral,grushin2016inhomogeneous,jia2019observation}.

According to this equation, the energy levels 
beyond $k_y=0$
are dispersive in the $k_{y}$ direction. When $n=0$, corresponding to zeroth order Landau level, it has a linear dispersion and the group velocities are opposite at lower and higher frequency. As for the higher order Landau levels, they are quantized with certain increments. In addition, we use the supercell shown in FIG.~\ref{fig:fig2}(a) to calculate the projected band structure. In FIG.~\ref{fig:fig2}(c), we show the excellent agreement of the zeroth Landau level and higher order Landau levels between numerical calculation and theoretical derivation in the vicinity of the $\bm{K}$ valley. There appear some slight differences when $\bm{k}$ is away from the $\bm{K}$ valley because the theoretical dispersion relation is approximated based on the approximate continuum Hamiltonian in the $\bm{K}$ valley. Besides, when $n$ becomes larger~(higher order Landau levels), the difference between numerical calculation and theoretical result becomes notable.

The eigenmodes of the zeroth Landau level are illustrated in FIG.~\ref{fig:fig2}(d), corresponding to red and blue dots at lower and higher frequency. The three degrees of freedom of our origami metamaterial $u_{t}$, $\phi_{t}$ and $\theta_{b}$ are concentrated near the middle of the supercell~($21^{st}$ unit cell). However, we also notice that compared with the distribution of confined states arising from topological states or defect states, the confined states in the zeroth Landau level are less concentrated.

\begin{figure}[h]
    \includegraphics[width=0.5\textwidth]{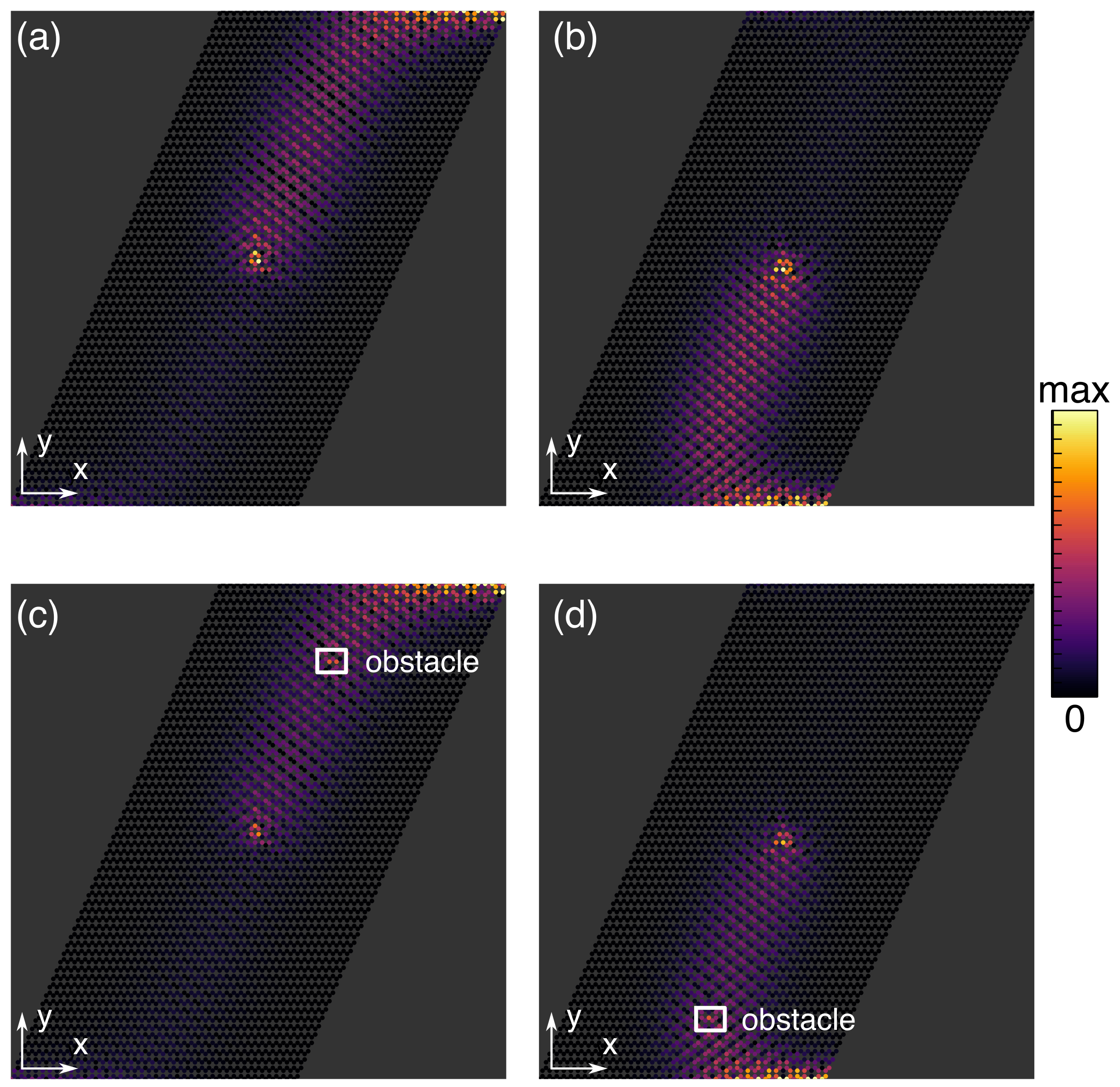}
    \caption{Robust zeroth Landau level.
    The field distributions $\theta_{b}$ for the zeroth Landau level in (a) lower frequency and (b) higher frequency. The corresponding cases with obstacles along the propagation path are shown in (c) and (d). Two excitation sources with phase difference are put in the middle. The excitation frequencies are $106~\mathrm{Hz}$ and $225~\mathrm{Hz}$, respectively.
    \label{fig:fig3}}
\end{figure}

We construct our origami metamaterial by extending the supercell along the direction of the wave vector so that an origami metamaterial with $60\times 41$ unit cells can be formed. We set the chiral excitation source on the axial displacement $u_{t}$ in the middle of the origami metamaterial to excite the modes subjected to the $\bm{K}$ valley corresponding to the zeroth Landau level. The corresponding frequency response of our origami metamaterials is shown in FIG.~\ref{fig:fig3}. FIG.~\ref{fig:fig3}(a) shows the eigenmode associated with the field distribution of $\theta_{b}$ under the excitation of the lower frequency~($106~\mathrm{Hz}$), where elastic waves can propagate along the $+y$ direction. In stark contrast, elastic waves can also propagate along the opposite direction under the excitation of the higher frequency~($225~\mathrm{Hz}$). This is due to the opposite group velocity of the zeroth Landau level in the two different frequency regions. The simulations of elastic wave propagation under corresponding frequencies are shown in the Supplemental Materials.

Although the Landau levels are considered as  bulk states, which are distinct from the topological edge states, the zeroth Landau level is topologically protected~\cite{jia2022experimental,jia2019observation}. This is on account of the negligible intervalley scattering~($\bm{K}$ and $\bm{K}'$ valleys are widely separated in $\bm{k}$ space), resulting in the weak backscattering of elastic waves. To confirm the robustness of the zeroth Landau level, obstacles~(origamis with double mass $m$ and rotational inertia $j$) are put along the wave propagation path while the same excitation is conducted. As shown in FIGs.~\ref{fig:fig3}(c)-(d), the elastic waves feature a linear response along the middle path similar to what is shown in FIGs.~\ref{fig:fig3}(a)-(b), instead of jamming and backscattering. This clearly exhibits the robustness of the transport of the chiral zeroth Landau level. We also notice in FIGs.~\ref{fig:fig3}(a)-(d) the elastic wave propagation along the top or bottom edge of the origami metamaterial, which may result from the boundary modes due to the free boundary conditions on the two edges subjected to the pseudomagnetic field.

\textit{Elastic snake states.}--The snake state is a novel transport phenomenon that has been observed in two-dimensional electron gases in $p$-$n$ junctions of graphene~\cite{taychatanapat2015conductance,oroszlany2008theory,ghosh2008conductance}, when subjected to an external magnetic field. This transport is characterized by a snake-like propagation path, which is driven by the Lorentz force due to the cyclotron motion of electrons in opposite magnetic fields at the interface of two domains with opposite magnetic field~\cite{muller1992effect}. It has been predicted that the pseudomagnetic field can also induce the snake states. In what follows the relation between the zeroth and first Landau levels in the projected band structure shown in FIG.~\ref{fig:fig2}(c) is exploited to realize elastic snake states in the origami metamaterial. The difference of topological phase on two domains provides two opposite topological charges, which makes snake states possible in our system with uniform pseudomagnetic field.

\begin{figure}[h]
    \includegraphics[width=0.5\textwidth]{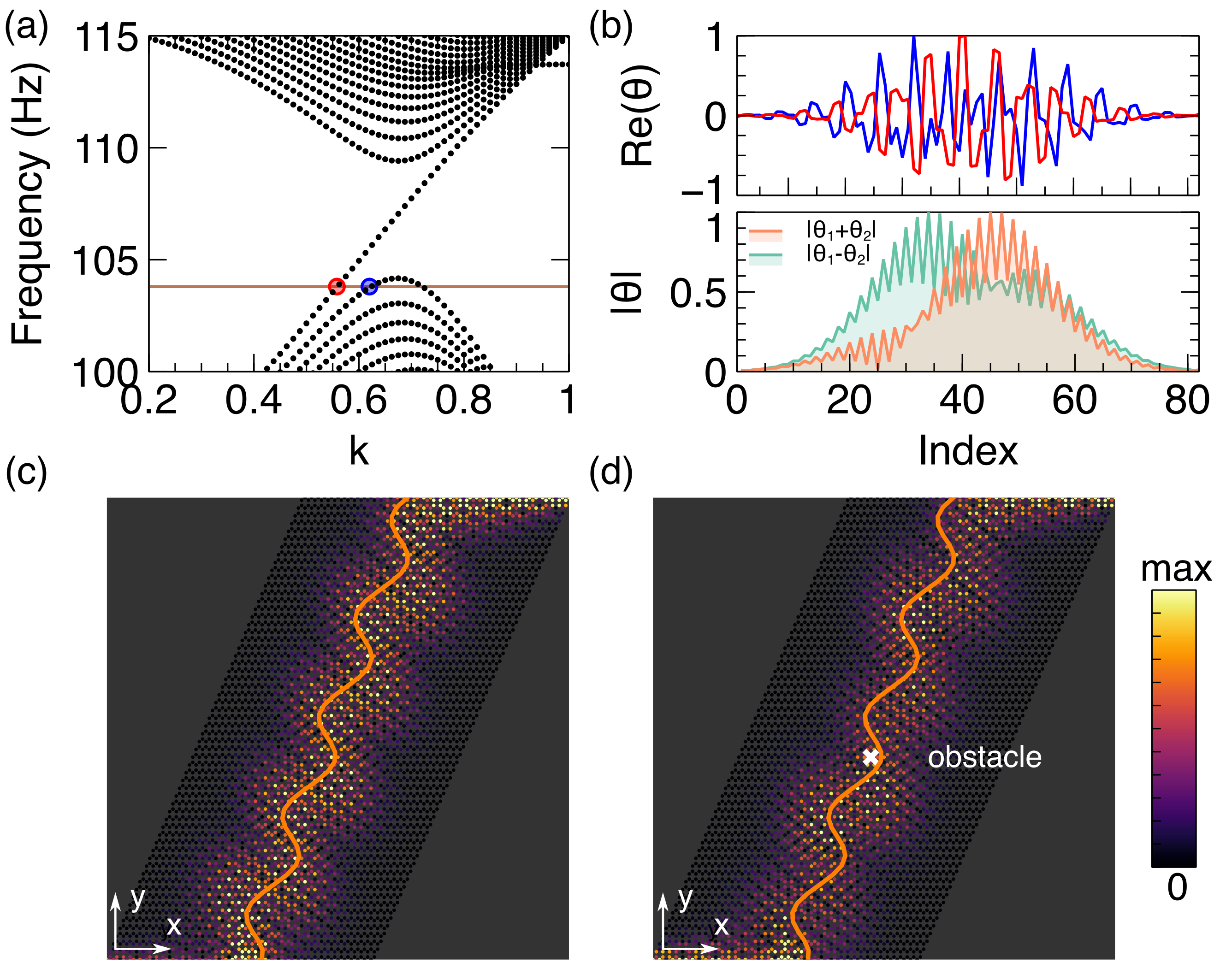}
    \caption{Elastic snake states.
    (a) The projected band structure of the supercell at the lower frequency.
    (b) Top panel: real parts of the eigenmode $\theta_{b1}$ and $\theta_{b2}$ along the supercell, corresponding to the red and blue dots in (a). Bottom panel: the newly-defined modes $\lvert\theta_{b1}+\theta_{b2}\rvert$ and $\lvert\theta_{b1}-\theta_{b2}\rvert$.
    (c) The simulated $\theta_{b}$ distribution under the excitation frequency at $103.9~\mathrm{Hz}$, corresponding to the solid line in (a).
    (d) The simulated $\theta$ distribution with the obstacle in the middle under the excitation frequency at $103.9~\mathrm{Hz}$.
    The theoretical trajectory of the snake state is shown in orange solid line in (c) and (d).
    \label{fig:fig4}}
\end{figure}

FIG.~\ref{fig:fig4}(a) shows the projected band structure at the lower frequency. Due to the square root relation in the dispersion when $n\neq0$~(higher order Landau level), there is a hump near the zeroth Landau level. We examine the states represented by $\theta_{b}$ of the zeroth and the first Landau levels with the positive group velocity, which are marked by red~($\theta_{b1}$ at $k_{1}$) and blue dots~($\theta_{b2}$ at $k_{2}$) in FIG.~\ref{fig:fig4}(a). These two states have definite parity with respect to mirror symmetry along the $x$ direction. The real parts of the field distributions in the top panel of FIG.~\ref{fig:fig4}(b) reveal that the states $\theta_{b1}$ and $\theta_{b2}$ have odd and even parities, respectively, suggesting they are orthogonal. In this way, we can define two other orthogonal states: $\langle+|=(\theta_{b1}+\theta_{b2})/\sqrt{2}$ and $\langle-|=(\theta_{b1}-\theta_{b2})/\sqrt{2}$. The state $\langle+|$ tends to be centered to the right of the middle~(of the domain), whereas the state $\langle-|$ tends to be centered to the left side of the middle~[bottom panel of FIG.~\ref{fig:fig4}(b)]. We can then derive $\theta_{b1}=(\langle+|+\langle-|)/\sqrt{2}$ and $\theta_{b2}=(\langle+|-\langle-|)/\sqrt{2}$. Using this basis, a general interface state $\psi$ propagating along the middle of the origami metamaterial can be expressed in the form $\psi=c_{1}\theta_{b1}e^{ik_{1}y}+c_{2}\theta_{b2}e^{ik_{2}y}$, where $c_{1}$ and $c_{2}$ are determined by the specific excitation. Assuming that the excitation is placed on the right side of the middle, i.e., $\psi(y=0)=\langle+|$, we have $c_{1}=c_{2}=1/\sqrt{2}$. Therefore, $|\langle+|\psi\rangle|^2=\cos^2(\delta k_{y}y)$ and $|\langle-|\psi\rangle|^2=\sin^2(\delta k_{y}y)$, where $\delta k_{y}=(k_{2}-k_{1})/2$. This implies that on the right side of the middle~($>20\sqrt{3}a$), the excited state will be oscillating in the form of $\cos^2(\delta k_{y}y)$. In contrast, on the left side of the middle~($<20\sqrt{3}a$), the excited state will be oscillating in the form of $\sin^2(\delta k_{y}y)$. Consequently, the excited state will propagate similar to a snake along the middle~($=20\sqrt{3}a$). The theoretical trajectory of the snake state is shown in orange solid lines in FIGs.~\ref{fig:fig4}(c) and (d), which has a good agreement with the simulation results.

In FIG.~\ref{fig:fig4}(c), we illustrate the elastic wave propagation depicting the snake state represented by $\theta_{b}$. The excitation source on the axial displacement $u_{t}$ with $103.9~\mathrm{Hz}$~[solid line in FIG.~\ref{fig:fig4}(a)] is put at the middle bottom of the origami metamaterial. The elastic wave starts with a clockwise rotation at the left side and then enters the right side with a counterclockwise rotation. This motion repeats, and the clear observation of the elastic snake state is illustrated. The time-dependent simulation of elastic wave propagation is shown in Supplemental Materials, where the wavy trajectory is demonstrated. Since the snake state originates from the bulk and thus does not have robustness, there will be strong backscattering when the elastic waves encounter the obstacles. Nevertheless, if the obstacles are placed at the appropriate positions along the middle of the origami metamaterial, the wavy snake state can bypass the obstacles, making it appear as if the objects are cloaked~[FIG.~\ref{fig:fig4}(d)].

\textit{Conclusions} \& \textit{Future challenges.}--In conclusion, our study presents a prototypical realization of an out-of-plane synthetic gauge field in origami metamaterials, leading to the emergence of a pseudomagnetic field that enables the quantization of Landau levels. The results of the numerical simulations and theoretical analysis are in good agreement, in the appropriate wavenumber range, confirming the validity of the long wavelength approximation used. By leveraging the unique dispersion relation of Landau levels, we have explored the topologically protected chiral elastic wave propagation and the emergence of elastic snake states. Therein, this metamaterial can serve as an effective alternative to topological elastic metamaterials due to its robust zeroth Landau level. The reconfigurability and simple manufacturing of origami structures make them attractive for a variety of possible applications such as waveguiding and obstacle bypassing due to the robust nature and associated wave propagation. Although the origami metamaterial in our study is in the centimeter scale, the scalability of this structure allows it to be applied to manipulate elastic waves at different frequencies. Besides, our scheme is not limited to Kresling origami, and other types of origami can be explored in the future. Importantly, the features considered herein have been limited to quasi-continuum linear features, in the vicinity of Dirac cones. The configuration, however, proposed herein bears a wide range of interesting phenomena including ones beyond the quasi-continuum approximation, as well as, importantly, nonlinear features within the band gaps. Studies along these veins are currently in progress and will be reported in future publications.

\noindent \textit{Acknowledgments}.--This material is based upon work supported by the U.S. National Science Foundation under Grant No. 1933729 and 2201612~(S.L. and J.K.), the awards PHY-2110030 and DMS-2204702~(P.G.K.).

\bibliography{references.bib}

\providecommand{\noopsort}[1]{}\providecommand{\singleletter}[1]{#1}%
\begin{thebibliography}{44}%
\makeatletter
\providecommand \@ifxundefined [1]{%
 \@ifx{#1\undefined}
}%
\providecommand \@ifnum [1]{%
 \ifnum #1\expandafter \@firstoftwo
 \else \expandafter \@secondoftwo
 \fi
}%
\providecommand \@ifx [1]{%
 \ifx #1\expandafter \@firstoftwo
 \else \expandafter \@secondoftwo
 \fi
}%
\providecommand \natexlab [1]{#1}%
\providecommand \enquote  [1]{``#1''}%
\providecommand \bibnamefont  [1]{#1}%
\providecommand \bibfnamefont [1]{#1}%
\providecommand \citenamefont [1]{#1}%
\providecommand \href@noop [0]{\@secondoftwo}%
\providecommand \href [0]{\begingroup \@sanitize@url \@href}%
\providecommand \@href[1]{\@@startlink{#1}\@@href}%
\providecommand \@@href[1]{\endgroup#1\@@endlink}%
\providecommand \@sanitize@url [0]{\catcode `\\12\catcode `\$12\catcode
  `\&12\catcode `\#12\catcode `\^12\catcode `\_12\catcode `\%12\relax}%
\providecommand \@@startlink[1]{}%
\providecommand \@@endlink[0]{}%
\providecommand \url  [0]{\begingroup\@sanitize@url \@url }%
\providecommand \@url [1]{\endgroup\@href {#1}{\urlprefix }}%
\providecommand \urlprefix  [0]{URL }%
\providecommand \Eprint [0]{\href }%
\providecommand \doibase [0]{https://doi.org/}%
\providecommand \selectlanguage [0]{\@gobble}%
\providecommand \bibinfo  [0]{\@secondoftwo}%
\providecommand \bibfield  [0]{\@secondoftwo}%
\providecommand \translation [1]{[#1]}%
\providecommand \BibitemOpen [0]{}%
\providecommand \bibitemStop [0]{}%
\providecommand \bibitemNoStop [0]{.\EOS\space}%
\providecommand \EOS [0]{\spacefactor3000\relax}%
\providecommand \BibitemShut  [1]{\csname bibitem#1\endcsname}%
\let\auto@bib@innerbib\@empty
\bibitem [{\citenamefont {Cooper}\ \emph {et~al.}(2019)\citenamefont {Cooper},
  \citenamefont {Dalibard},\ and\ \citenamefont
  {Spielman}}]{cooper2019topological}%
  \BibitemOpen
  \bibfield  {author} {\bibinfo {author} {\bibfnamefont {N.}~\bibnamefont
  {Cooper}}, \bibinfo {author} {\bibfnamefont {J.}~\bibnamefont {Dalibard}},\
  and\ \bibinfo {author} {\bibfnamefont {I.}~\bibnamefont {Spielman}},\
  }\bibfield  {title} {\bibinfo {title} {Topological bands for ultracold
  atoms},\ }\href@noop {} {\bibfield  {journal} {\bibinfo  {journal} {Reviews
  of modern physics}\ }\textbf {\bibinfo {volume} {91}},\ \bibinfo {pages}
  {015005} (\bibinfo {year} {2019})}\BibitemShut {NoStop}%
\bibitem [{\citenamefont {Fang}\ and\ \citenamefont
  {Fan}(2013)}]{fang2013controlling}%
  \BibitemOpen
  \bibfield  {author} {\bibinfo {author} {\bibfnamefont {K.}~\bibnamefont
  {Fang}}\ and\ \bibinfo {author} {\bibfnamefont {S.}~\bibnamefont {Fan}},\
  }\bibfield  {title} {\bibinfo {title} {Controlling the flow of light using
  the inhomogeneous effective gauge field that emerges from dynamic
  modulation},\ }\href@noop {} {\bibfield  {journal} {\bibinfo  {journal}
  {Physical review letters}\ }\textbf {\bibinfo {volume} {111}},\ \bibinfo
  {pages} {203901} (\bibinfo {year} {2013})}\BibitemShut {NoStop}%
\bibitem [{\citenamefont {Lin}\ and\ \citenamefont {Fan}(2014)}]{lin2014light}%
  \BibitemOpen
  \bibfield  {author} {\bibinfo {author} {\bibfnamefont {Q.}~\bibnamefont
  {Lin}}\ and\ \bibinfo {author} {\bibfnamefont {S.}~\bibnamefont {Fan}},\
  }\bibfield  {title} {\bibinfo {title} {Light guiding by effective gauge field
  for photons},\ }\href@noop {} {\bibfield  {journal} {\bibinfo  {journal}
  {Physical Review X}\ }\textbf {\bibinfo {volume} {4}},\ \bibinfo {pages}
  {031031} (\bibinfo {year} {2014})}\BibitemShut {NoStop}%
\bibitem [{\citenamefont {Yang}\ \emph {et~al.}(2021)\citenamefont {Yang},
  \citenamefont {Ge}, \citenamefont {Li}, \citenamefont {Lin}, \citenamefont
  {Jia}, \citenamefont {Guan}, \citenamefont {Yuan}, \citenamefont {Sun},
  \citenamefont {Chong},\ and\ \citenamefont {Zhang}}]{yang2021demonstration}%
  \BibitemOpen
  \bibfield  {author} {\bibinfo {author} {\bibfnamefont {Y.}~\bibnamefont
  {Yang}}, \bibinfo {author} {\bibfnamefont {Y.}~\bibnamefont {Ge}}, \bibinfo
  {author} {\bibfnamefont {R.}~\bibnamefont {Li}}, \bibinfo {author}
  {\bibfnamefont {X.}~\bibnamefont {Lin}}, \bibinfo {author} {\bibfnamefont
  {D.}~\bibnamefont {Jia}}, \bibinfo {author} {\bibfnamefont {Y.-j.}\
  \bibnamefont {Guan}}, \bibinfo {author} {\bibfnamefont {S.-q.}\ \bibnamefont
  {Yuan}}, \bibinfo {author} {\bibfnamefont {H.-x.}\ \bibnamefont {Sun}},
  \bibinfo {author} {\bibfnamefont {Y.}~\bibnamefont {Chong}},\ and\ \bibinfo
  {author} {\bibfnamefont {B.}~\bibnamefont {Zhang}},\ }\bibfield  {title}
  {\bibinfo {title} {Demonstration of negative refraction induced by synthetic
  gauge fields},\ }\href@noop {} {\bibfield  {journal} {\bibinfo  {journal}
  {Science Advances}\ }\textbf {\bibinfo {volume} {7}},\ \bibinfo {pages}
  {eabj2062} (\bibinfo {year} {2021})}\BibitemShut {NoStop}%
\bibitem [{\citenamefont {Lu}\ \emph {et~al.}(2014{\natexlab{a}})\citenamefont
  {Lu}, \citenamefont {Joannopoulos},\ and\ \citenamefont
  {Solja{\v{c}}i{\'c}}}]{lu2014topological}%
  \BibitemOpen
  \bibfield  {author} {\bibinfo {author} {\bibfnamefont {L.}~\bibnamefont
  {Lu}}, \bibinfo {author} {\bibfnamefont {J.~D.}\ \bibnamefont
  {Joannopoulos}},\ and\ \bibinfo {author} {\bibfnamefont {M.}~\bibnamefont
  {Solja{\v{c}}i{\'c}}},\ }\bibfield  {title} {\bibinfo {title} {Topological
  photonics},\ }\href@noop {} {\bibfield  {journal} {\bibinfo  {journal}
  {Nature photonics}\ }\textbf {\bibinfo {volume} {8}},\ \bibinfo {pages} {821}
  (\bibinfo {year} {2014}{\natexlab{a}})}\BibitemShut {NoStop}%
\bibitem [{\citenamefont {Khanikaev}\ and\ \citenamefont
  {Shvets}(2017)}]{khanikaev2017two}%
  \BibitemOpen
  \bibfield  {author} {\bibinfo {author} {\bibfnamefont {A.~B.}\ \bibnamefont
  {Khanikaev}}\ and\ \bibinfo {author} {\bibfnamefont {G.}~\bibnamefont
  {Shvets}},\ }\bibfield  {title} {\bibinfo {title} {Two-dimensional
  topological photonics},\ }\href@noop {} {\bibfield  {journal} {\bibinfo
  {journal} {Nature photonics}\ }\textbf {\bibinfo {volume} {11}},\ \bibinfo
  {pages} {763} (\bibinfo {year} {2017})}\BibitemShut {NoStop}%
\bibitem [{\citenamefont {Yang}\ \emph {et~al.}(2015)\citenamefont {Yang},
  \citenamefont {Gao}, \citenamefont {Shi}, \citenamefont {Lin}, \citenamefont
  {Gao}, \citenamefont {Chong},\ and\ \citenamefont
  {Zhang}}]{yang2015topological}%
  \BibitemOpen
  \bibfield  {author} {\bibinfo {author} {\bibfnamefont {Z.}~\bibnamefont
  {Yang}}, \bibinfo {author} {\bibfnamefont {F.}~\bibnamefont {Gao}}, \bibinfo
  {author} {\bibfnamefont {X.}~\bibnamefont {Shi}}, \bibinfo {author}
  {\bibfnamefont {X.}~\bibnamefont {Lin}}, \bibinfo {author} {\bibfnamefont
  {Z.}~\bibnamefont {Gao}}, \bibinfo {author} {\bibfnamefont {Y.}~\bibnamefont
  {Chong}},\ and\ \bibinfo {author} {\bibfnamefont {B.}~\bibnamefont {Zhang}},\
  }\bibfield  {title} {\bibinfo {title} {Topological acoustics},\ }\href@noop
  {} {\bibfield  {journal} {\bibinfo  {journal} {Physical review letters}\
  }\textbf {\bibinfo {volume} {114}},\ \bibinfo {pages} {114301} (\bibinfo
  {year} {2015})}\BibitemShut {NoStop}%
\bibitem [{\citenamefont {Abbaszadeh}\ \emph {et~al.}(2017)\citenamefont
  {Abbaszadeh}, \citenamefont {Souslov}, \citenamefont {Paulose}, \citenamefont
  {Schomerus},\ and\ \citenamefont {Vitelli}}]{abbaszadeh2017sonic}%
  \BibitemOpen
  \bibfield  {author} {\bibinfo {author} {\bibfnamefont {H.}~\bibnamefont
  {Abbaszadeh}}, \bibinfo {author} {\bibfnamefont {A.}~\bibnamefont {Souslov}},
  \bibinfo {author} {\bibfnamefont {J.}~\bibnamefont {Paulose}}, \bibinfo
  {author} {\bibfnamefont {H.}~\bibnamefont {Schomerus}},\ and\ \bibinfo
  {author} {\bibfnamefont {V.}~\bibnamefont {Vitelli}},\ }\bibfield  {title}
  {\bibinfo {title} {Sonic landau levels and synthetic gauge fields in
  mechanical metamaterials},\ }\href@noop {} {\bibfield  {journal} {\bibinfo
  {journal} {Physical review letters}\ }\textbf {\bibinfo {volume} {119}},\
  \bibinfo {pages} {195502} (\bibinfo {year} {2017})}\BibitemShut {NoStop}%
\bibitem [{\citenamefont {Peri}\ \emph {et~al.}(2019)\citenamefont {Peri},
  \citenamefont {Serra-Garcia}, \citenamefont {Ilan},\ and\ \citenamefont
  {Huber}}]{peri2019axial}%
  \BibitemOpen
  \bibfield  {author} {\bibinfo {author} {\bibfnamefont {V.}~\bibnamefont
  {Peri}}, \bibinfo {author} {\bibfnamefont {M.}~\bibnamefont {Serra-Garcia}},
  \bibinfo {author} {\bibfnamefont {R.}~\bibnamefont {Ilan}},\ and\ \bibinfo
  {author} {\bibfnamefont {S.~D.}\ \bibnamefont {Huber}},\ }\bibfield  {title}
  {\bibinfo {title} {Axial-field-induced chiral channels in an acoustic weyl
  system},\ }\href@noop {} {\bibfield  {journal} {\bibinfo  {journal} {Nature
  Physics}\ }\textbf {\bibinfo {volume} {15}},\ \bibinfo {pages} {357}
  (\bibinfo {year} {2019})}\BibitemShut {NoStop}%
\bibitem [{\citenamefont {Yang}\ \emph {et~al.}(2017)\citenamefont {Yang},
  \citenamefont {Gao}, \citenamefont {Yang},\ and\ \citenamefont
  {Zhang}}]{yang2017strain}%
  \BibitemOpen
  \bibfield  {author} {\bibinfo {author} {\bibfnamefont {Z.}~\bibnamefont
  {Yang}}, \bibinfo {author} {\bibfnamefont {F.}~\bibnamefont {Gao}}, \bibinfo
  {author} {\bibfnamefont {Y.}~\bibnamefont {Yang}},\ and\ \bibinfo {author}
  {\bibfnamefont {B.}~\bibnamefont {Zhang}},\ }\bibfield  {title} {\bibinfo
  {title} {Strain-induced gauge field and landau levels in acoustic
  structures},\ }\href@noop {} {\bibfield  {journal} {\bibinfo  {journal}
  {Physical Review Letters}\ }\textbf {\bibinfo {volume} {118}},\ \bibinfo
  {pages} {194301} (\bibinfo {year} {2017})}\BibitemShut {NoStop}%
\bibitem [{\citenamefont {Brendel}\ \emph {et~al.}(2017)\citenamefont
  {Brendel}, \citenamefont {Peano}, \citenamefont {Painter},\ and\
  \citenamefont {Marquardt}}]{brendel2017pseudomagnetic}%
  \BibitemOpen
  \bibfield  {author} {\bibinfo {author} {\bibfnamefont {C.}~\bibnamefont
  {Brendel}}, \bibinfo {author} {\bibfnamefont {V.}~\bibnamefont {Peano}},
  \bibinfo {author} {\bibfnamefont {O.~J.}\ \bibnamefont {Painter}},\ and\
  \bibinfo {author} {\bibfnamefont {F.}~\bibnamefont {Marquardt}},\ }\bibfield
  {title} {\bibinfo {title} {Pseudomagnetic fields for sound at the
  nanoscale},\ }\href@noop {} {\bibfield  {journal} {\bibinfo  {journal}
  {Proceedings of the National Academy of Sciences}\ }\textbf {\bibinfo
  {volume} {114}},\ \bibinfo {pages} {E3390} (\bibinfo {year}
  {2017})}\BibitemShut {NoStop}%
\bibitem [{\citenamefont {Wen}\ \emph {et~al.}(2019)\citenamefont {Wen},
  \citenamefont {Qiu}, \citenamefont {Qi}, \citenamefont {Ye}, \citenamefont
  {Ke}, \citenamefont {Zhang},\ and\ \citenamefont {Liu}}]{wen2019acoustic}%
  \BibitemOpen
  \bibfield  {author} {\bibinfo {author} {\bibfnamefont {X.}~\bibnamefont
  {Wen}}, \bibinfo {author} {\bibfnamefont {C.}~\bibnamefont {Qiu}}, \bibinfo
  {author} {\bibfnamefont {Y.}~\bibnamefont {Qi}}, \bibinfo {author}
  {\bibfnamefont {L.}~\bibnamefont {Ye}}, \bibinfo {author} {\bibfnamefont
  {M.}~\bibnamefont {Ke}}, \bibinfo {author} {\bibfnamefont {F.}~\bibnamefont
  {Zhang}},\ and\ \bibinfo {author} {\bibfnamefont {Z.}~\bibnamefont {Liu}},\
  }\bibfield  {title} {\bibinfo {title} {Acoustic landau quantization and
  quantum-hall-like edge states},\ }\href@noop {} {\bibfield  {journal}
  {\bibinfo  {journal} {Nature Physics}\ }\textbf {\bibinfo {volume} {15}},\
  \bibinfo {pages} {352} (\bibinfo {year} {2019})}\BibitemShut {NoStop}%
\bibitem [{\citenamefont {Yan}\ \emph {et~al.}(2021)\citenamefont {Yan},
  \citenamefont {Deng}, \citenamefont {Huang}, \citenamefont {Wu},
  \citenamefont {Yang}, \citenamefont {Lu}, \citenamefont {Li},\ and\
  \citenamefont {Liu}}]{yan2021pseudomagnetic}%
  \BibitemOpen
  \bibfield  {author} {\bibinfo {author} {\bibfnamefont {M.}~\bibnamefont
  {Yan}}, \bibinfo {author} {\bibfnamefont {W.}~\bibnamefont {Deng}}, \bibinfo
  {author} {\bibfnamefont {X.}~\bibnamefont {Huang}}, \bibinfo {author}
  {\bibfnamefont {Y.}~\bibnamefont {Wu}}, \bibinfo {author} {\bibfnamefont
  {Y.}~\bibnamefont {Yang}}, \bibinfo {author} {\bibfnamefont {J.}~\bibnamefont
  {Lu}}, \bibinfo {author} {\bibfnamefont {F.}~\bibnamefont {Li}},\ and\
  \bibinfo {author} {\bibfnamefont {Z.}~\bibnamefont {Liu}},\ }\bibfield
  {title} {\bibinfo {title} {Pseudomagnetic fields enabled manipulation of
  on-chip elastic waves},\ }\href@noop {} {\bibfield  {journal} {\bibinfo
  {journal} {Physical Review Letters}\ }\textbf {\bibinfo {volume} {127}},\
  \bibinfo {pages} {136401} (\bibinfo {year} {2021})}\BibitemShut {NoStop}%
\bibitem [{\citenamefont {Lv}\ \emph {et~al.}(2014)\citenamefont {Lv},
  \citenamefont {Krishnaraju}, \citenamefont {Konjevod}, \citenamefont {Yu},\
  and\ \citenamefont {Jiang}}]{lv2014origami}%
  \BibitemOpen
  \bibfield  {author} {\bibinfo {author} {\bibfnamefont {C.}~\bibnamefont
  {Lv}}, \bibinfo {author} {\bibfnamefont {D.}~\bibnamefont {Krishnaraju}},
  \bibinfo {author} {\bibfnamefont {G.}~\bibnamefont {Konjevod}}, \bibinfo
  {author} {\bibfnamefont {H.}~\bibnamefont {Yu}},\ and\ \bibinfo {author}
  {\bibfnamefont {H.}~\bibnamefont {Jiang}},\ }\bibfield  {title} {\bibinfo
  {title} {Origami based mechanical metamaterials},\ }\href@noop {} {\bibfield
  {journal} {\bibinfo  {journal} {Scientific reports}\ }\textbf {\bibinfo
  {volume} {4}},\ \bibinfo {pages} {1} (\bibinfo {year} {2014})}\BibitemShut
  {NoStop}%
\bibitem [{\citenamefont {Filipov}\ \emph {et~al.}(2015)\citenamefont
  {Filipov}, \citenamefont {Tachi},\ and\ \citenamefont
  {Paulino}}]{filipov2015origami}%
  \BibitemOpen
  \bibfield  {author} {\bibinfo {author} {\bibfnamefont {E.~T.}\ \bibnamefont
  {Filipov}}, \bibinfo {author} {\bibfnamefont {T.}~\bibnamefont {Tachi}},\
  and\ \bibinfo {author} {\bibfnamefont {G.~H.}\ \bibnamefont {Paulino}},\
  }\bibfield  {title} {\bibinfo {title} {Origami tubes assembled into stiff,
  yet reconfigurable structures and metamaterials},\ }\href@noop {} {\bibfield
  {journal} {\bibinfo  {journal} {Proceedings of the National Academy of
  Sciences}\ }\textbf {\bibinfo {volume} {112}},\ \bibinfo {pages} {12321}
  (\bibinfo {year} {2015})}\BibitemShut {NoStop}%
\bibitem [{\citenamefont {Wickeler}\ and\ \citenamefont
  {Naguib}(2020)}]{wickeler2020novel}%
  \BibitemOpen
  \bibfield  {author} {\bibinfo {author} {\bibfnamefont {A.~L.}\ \bibnamefont
  {Wickeler}}\ and\ \bibinfo {author} {\bibfnamefont {H.~E.}\ \bibnamefont
  {Naguib}},\ }\bibfield  {title} {\bibinfo {title} {Novel origami-inspired
  metamaterials: Design, mechanical testing and finite element modelling},\
  }\href@noop {} {\bibfield  {journal} {\bibinfo  {journal} {Materials \&
  Design}\ }\textbf {\bibinfo {volume} {186}},\ \bibinfo {pages} {108242}
  (\bibinfo {year} {2020})}\BibitemShut {NoStop}%
\bibitem [{\citenamefont {Zirbel}\ \emph {et~al.}(2013)\citenamefont {Zirbel},
  \citenamefont {Lang}, \citenamefont {Thomson}, \citenamefont {Sigel},
  \citenamefont {Walkemeyer}, \citenamefont {Trease}, \citenamefont {Magleby},\
  and\ \citenamefont {Howell}}]{zirbel2013accommodating}%
  \BibitemOpen
  \bibfield  {author} {\bibinfo {author} {\bibfnamefont {S.~A.}\ \bibnamefont
  {Zirbel}}, \bibinfo {author} {\bibfnamefont {R.~J.}\ \bibnamefont {Lang}},
  \bibinfo {author} {\bibfnamefont {M.~W.}\ \bibnamefont {Thomson}}, \bibinfo
  {author} {\bibfnamefont {D.~A.}\ \bibnamefont {Sigel}}, \bibinfo {author}
  {\bibfnamefont {P.~E.}\ \bibnamefont {Walkemeyer}}, \bibinfo {author}
  {\bibfnamefont {B.~P.}\ \bibnamefont {Trease}}, \bibinfo {author}
  {\bibfnamefont {S.~P.}\ \bibnamefont {Magleby}},\ and\ \bibinfo {author}
  {\bibfnamefont {L.~L.}\ \bibnamefont {Howell}},\ }\bibfield  {title}
  {\bibinfo {title} {Accommodating thickness in origami-based deployable
  arrays},\ }\href@noop {} {\bibfield  {journal} {\bibinfo  {journal} {Journal
  of Mechanical Design}\ }\textbf {\bibinfo {volume} {135}} (\bibinfo {year}
  {2013})}\BibitemShut {NoStop}%
\bibitem [{\citenamefont {Del~Grosso}\ and\ \citenamefont
  {Basso}(2010)}]{del2010adaptive}%
  \BibitemOpen
  \bibfield  {author} {\bibinfo {author} {\bibfnamefont {A.~E.}\ \bibnamefont
  {Del~Grosso}}\ and\ \bibinfo {author} {\bibfnamefont {P.}~\bibnamefont
  {Basso}},\ }\bibfield  {title} {\bibinfo {title} {Adaptive building skin
  structures},\ }\href@noop {} {\bibfield  {journal} {\bibinfo  {journal}
  {Smart Materials and Structures}\ }\textbf {\bibinfo {volume} {19}},\
  \bibinfo {pages} {124011} (\bibinfo {year} {2010})}\BibitemShut {NoStop}%
\bibitem [{\citenamefont {Lyu}\ \emph {et~al.}(2021)\citenamefont {Lyu},
  \citenamefont {Qin}, \citenamefont {Deng},\ and\ \citenamefont
  {Ding}}]{lyu2021origami}%
  \BibitemOpen
  \bibfield  {author} {\bibinfo {author} {\bibfnamefont {S.}~\bibnamefont
  {Lyu}}, \bibinfo {author} {\bibfnamefont {B.}~\bibnamefont {Qin}}, \bibinfo
  {author} {\bibfnamefont {H.}~\bibnamefont {Deng}},\ and\ \bibinfo {author}
  {\bibfnamefont {X.}~\bibnamefont {Ding}},\ }\bibfield  {title} {\bibinfo
  {title} {Origami-based cellular mechanical metamaterials with tunable
  poisson's ratio: Construction and analysis},\ }\href@noop {} {\bibfield
  {journal} {\bibinfo  {journal} {International Journal of Mechanical
  Sciences}\ }\textbf {\bibinfo {volume} {212}},\ \bibinfo {pages} {106791}
  (\bibinfo {year} {2021})}\BibitemShut {NoStop}%
\bibitem [{\citenamefont {Miyashita}\ \emph {et~al.}(2015)\citenamefont
  {Miyashita}, \citenamefont {Guitron}, \citenamefont {Ludersdorfer},
  \citenamefont {Sung},\ and\ \citenamefont {Rus}}]{miyashita2015untethered}%
  \BibitemOpen
  \bibfield  {author} {\bibinfo {author} {\bibfnamefont {S.}~\bibnamefont
  {Miyashita}}, \bibinfo {author} {\bibfnamefont {S.}~\bibnamefont {Guitron}},
  \bibinfo {author} {\bibfnamefont {M.}~\bibnamefont {Ludersdorfer}}, \bibinfo
  {author} {\bibfnamefont {C.~R.}\ \bibnamefont {Sung}},\ and\ \bibinfo
  {author} {\bibfnamefont {D.}~\bibnamefont {Rus}},\ }\bibfield  {title}
  {\bibinfo {title} {An untethered miniature origami robot that self-folds,
  walks, swims, and degrades},\ }in\ \href@noop {} {\emph {\bibinfo {booktitle}
  {2015 IEEE International Conference on Robotics and Automation (ICRA)}}}\
  (\bibinfo {organization} {IEEE},\ \bibinfo {year} {2015})\ pp.\ \bibinfo
  {pages} {1490--1496}\BibitemShut {NoStop}%
\bibitem [{\citenamefont {Edmondson}\ \emph {et~al.}(2013)\citenamefont
  {Edmondson}, \citenamefont {Bowen}, \citenamefont {Grames}, \citenamefont
  {Magleby}, \citenamefont {Howell},\ and\ \citenamefont
  {Bateman}}]{edmondson2013oriceps}%
  \BibitemOpen
  \bibfield  {author} {\bibinfo {author} {\bibfnamefont {B.~J.}\ \bibnamefont
  {Edmondson}}, \bibinfo {author} {\bibfnamefont {L.~A.}\ \bibnamefont
  {Bowen}}, \bibinfo {author} {\bibfnamefont {C.~L.}\ \bibnamefont {Grames}},
  \bibinfo {author} {\bibfnamefont {S.~P.}\ \bibnamefont {Magleby}}, \bibinfo
  {author} {\bibfnamefont {L.~L.}\ \bibnamefont {Howell}},\ and\ \bibinfo
  {author} {\bibfnamefont {T.~C.}\ \bibnamefont {Bateman}},\ }\bibfield
  {title} {\bibinfo {title} {Oriceps: Origami-inspired forceps},\ }in\
  \href@noop {} {\emph {\bibinfo {booktitle} {Smart Materials, Adaptive
  Structures and Intelligent Systems}}},\ Vol.\ \bibinfo {volume} {56031}\
  (\bibinfo {organization} {American Society of Mechanical Engineers},\
  \bibinfo {year} {2013})\ p.\ \bibinfo {pages} {V001T01A027}\BibitemShut
  {NoStop}%
\bibitem [{\citenamefont {Kuribayashi}\ \emph {et~al.}(2006)\citenamefont
  {Kuribayashi}, \citenamefont {Tsuchiya}, \citenamefont {You}, \citenamefont
  {Tomus}, \citenamefont {Umemoto}, \citenamefont {Ito},\ and\ \citenamefont
  {Sasaki}}]{kuribayashi2006self}%
  \BibitemOpen
  \bibfield  {author} {\bibinfo {author} {\bibfnamefont {K.}~\bibnamefont
  {Kuribayashi}}, \bibinfo {author} {\bibfnamefont {K.}~\bibnamefont
  {Tsuchiya}}, \bibinfo {author} {\bibfnamefont {Z.}~\bibnamefont {You}},
  \bibinfo {author} {\bibfnamefont {D.}~\bibnamefont {Tomus}}, \bibinfo
  {author} {\bibfnamefont {M.}~\bibnamefont {Umemoto}}, \bibinfo {author}
  {\bibfnamefont {T.}~\bibnamefont {Ito}},\ and\ \bibinfo {author}
  {\bibfnamefont {M.}~\bibnamefont {Sasaki}},\ }\bibfield  {title} {\bibinfo
  {title} {Self-deployable origami stent grafts as a biomedical application of
  ni-rich tini shape memory alloy foil},\ }\href@noop {} {\bibfield  {journal}
  {\bibinfo  {journal} {Materials Science and Engineering: A}\ }\textbf
  {\bibinfo {volume} {419}},\ \bibinfo {pages} {131} (\bibinfo {year}
  {2006})}\BibitemShut {NoStop}%
\bibitem [{\citenamefont {Zhou}\ \emph {et~al.}(2016)\citenamefont {Zhou},
  \citenamefont {Wang}, \citenamefont {Ma},\ and\ \citenamefont
  {You}}]{zhou2016dynamic}%
  \BibitemOpen
  \bibfield  {author} {\bibinfo {author} {\bibfnamefont {C.}~\bibnamefont
  {Zhou}}, \bibinfo {author} {\bibfnamefont {B.}~\bibnamefont {Wang}}, \bibinfo
  {author} {\bibfnamefont {J.}~\bibnamefont {Ma}},\ and\ \bibinfo {author}
  {\bibfnamefont {Z.}~\bibnamefont {You}},\ }\bibfield  {title} {\bibinfo
  {title} {Dynamic axial crushing of origami crash boxes},\ }\href@noop {}
  {\bibfield  {journal} {\bibinfo  {journal} {International journal of
  mechanical sciences}\ }\textbf {\bibinfo {volume} {118}},\ \bibinfo {pages}
  {1} (\bibinfo {year} {2016})}\BibitemShut {NoStop}%
\bibitem [{\citenamefont {Yasuda}\ \emph {et~al.}(2020)\citenamefont {Yasuda},
  \citenamefont {Yamaguchi}, \citenamefont {Miyazawa}, \citenamefont {Wiebe},
  \citenamefont {Raney},\ and\ \citenamefont {Yang}}]{yasuda2020data}%
  \BibitemOpen
  \bibfield  {author} {\bibinfo {author} {\bibfnamefont {H.}~\bibnamefont
  {Yasuda}}, \bibinfo {author} {\bibfnamefont {K.}~\bibnamefont {Yamaguchi}},
  \bibinfo {author} {\bibfnamefont {Y.}~\bibnamefont {Miyazawa}}, \bibinfo
  {author} {\bibfnamefont {R.}~\bibnamefont {Wiebe}}, \bibinfo {author}
  {\bibfnamefont {J.~R.}\ \bibnamefont {Raney}},\ and\ \bibinfo {author}
  {\bibfnamefont {J.}~\bibnamefont {Yang}},\ }\bibfield  {title} {\bibinfo
  {title} {Data-driven prediction and analysis of chaotic origami dynamics},\
  }\href@noop {} {\bibfield  {journal} {\bibinfo  {journal} {Communications
  Physics}\ }\textbf {\bibinfo {volume} {3}},\ \bibinfo {pages} {1} (\bibinfo
  {year} {2020})}\BibitemShut {NoStop}%
\bibitem [{\citenamefont {Yasuda}\ \emph {et~al.}(2019)\citenamefont {Yasuda},
  \citenamefont {Miyazawa}, \citenamefont {Charalampidis}, \citenamefont
  {Chong}, \citenamefont {Kevrekidis},\ and\ \citenamefont
  {Yang}}]{yasuda2019origami}%
  \BibitemOpen
  \bibfield  {author} {\bibinfo {author} {\bibfnamefont {H.}~\bibnamefont
  {Yasuda}}, \bibinfo {author} {\bibfnamefont {Y.}~\bibnamefont {Miyazawa}},
  \bibinfo {author} {\bibfnamefont {E.~G.}\ \bibnamefont {Charalampidis}},
  \bibinfo {author} {\bibfnamefont {C.}~\bibnamefont {Chong}}, \bibinfo
  {author} {\bibfnamefont {P.~G.}\ \bibnamefont {Kevrekidis}},\ and\ \bibinfo
  {author} {\bibfnamefont {J.}~\bibnamefont {Yang}},\ }\bibfield  {title}
  {\bibinfo {title} {Origami-based impact mitigation via rarefaction solitary
  wave creation},\ }\href@noop {} {\bibfield  {journal} {\bibinfo  {journal}
  {Science advances}\ }\textbf {\bibinfo {volume} {5}},\ \bibinfo {pages}
  {eaau2835} (\bibinfo {year} {2019})}\BibitemShut {NoStop}%
\bibitem [{\citenamefont {Miyazawa}\ \emph {et~al.}(2022)\citenamefont
  {Miyazawa}, \citenamefont {Chen}, \citenamefont {Chaunsali}, \citenamefont
  {Gormley}, \citenamefont {Yin}, \citenamefont {Theocharis},\ and\
  \citenamefont {Yang}}]{miyazawa2022topological}%
  \BibitemOpen
  \bibfield  {author} {\bibinfo {author} {\bibfnamefont {Y.}~\bibnamefont
  {Miyazawa}}, \bibinfo {author} {\bibfnamefont {C.-W.}\ \bibnamefont {Chen}},
  \bibinfo {author} {\bibfnamefont {R.}~\bibnamefont {Chaunsali}}, \bibinfo
  {author} {\bibfnamefont {T.~S.}\ \bibnamefont {Gormley}}, \bibinfo {author}
  {\bibfnamefont {G.}~\bibnamefont {Yin}}, \bibinfo {author} {\bibfnamefont
  {G.}~\bibnamefont {Theocharis}},\ and\ \bibinfo {author} {\bibfnamefont
  {J.}~\bibnamefont {Yang}},\ }\bibfield  {title} {\bibinfo {title}
  {Topological state transfer in kresling origami},\ }\href@noop {} {\bibfield
  {journal} {\bibinfo  {journal} {Communications Materials}\ }\textbf {\bibinfo
  {volume} {3}},\ \bibinfo {pages} {1} (\bibinfo {year} {2022})}\BibitemShut
  {NoStop}%
\bibitem [{\citenamefont {Chen}\ \emph {et~al.}(2016)\citenamefont {Chen},
  \citenamefont {Liu}, \citenamefont {Evans}, \citenamefont {Paulose},
  \citenamefont {Cohen}, \citenamefont {Vitelli},\ and\ \citenamefont
  {Santangelo}}]{chen2016topological}%
  \BibitemOpen
  \bibfield  {author} {\bibinfo {author} {\bibfnamefont {B.~G.-g.}\
  \bibnamefont {Chen}}, \bibinfo {author} {\bibfnamefont {B.}~\bibnamefont
  {Liu}}, \bibinfo {author} {\bibfnamefont {A.~A.}\ \bibnamefont {Evans}},
  \bibinfo {author} {\bibfnamefont {J.}~\bibnamefont {Paulose}}, \bibinfo
  {author} {\bibfnamefont {I.}~\bibnamefont {Cohen}}, \bibinfo {author}
  {\bibfnamefont {V.}~\bibnamefont {Vitelli}},\ and\ \bibinfo {author}
  {\bibfnamefont {C.}~\bibnamefont {Santangelo}},\ }\bibfield  {title}
  {\bibinfo {title} {Topological mechanics of origami and kirigami},\
  }\href@noop {} {\bibfield  {journal} {\bibinfo  {journal} {Physical review
  letters}\ }\textbf {\bibinfo {volume} {116}},\ \bibinfo {pages} {135501}
  (\bibinfo {year} {2016})}\BibitemShut {NoStop}%
\bibitem [{\citenamefont {Li}\ \emph {et~al.}(2023)\citenamefont {Li},
  \citenamefont {Miyazawa}, \citenamefont {Yamaguchi}, \citenamefont
  {Kevrekidis},\ and\ \citenamefont {Yang}}]{li2023origami}%
  \BibitemOpen
  \bibfield  {author} {\bibinfo {author} {\bibfnamefont {S.}~\bibnamefont
  {Li}}, \bibinfo {author} {\bibfnamefont {Y.}~\bibnamefont {Miyazawa}},
  \bibinfo {author} {\bibfnamefont {K.}~\bibnamefont {Yamaguchi}}, \bibinfo
  {author} {\bibfnamefont {P.~G.}\ \bibnamefont {Kevrekidis}},\ and\ \bibinfo
  {author} {\bibfnamefont {J.}~\bibnamefont {Yang}},\ }\href
  {https://doi.org/10.48550/ARXIV.2303.04323} {\bibinfo {title}
  {Geometry-informed dynamic mode decomposition in origami dynamics}} (\bibinfo
  {year} {2023})\BibitemShut {NoStop}%
\bibitem [{\citenamefont {Wu}\ \emph {et~al.}(2018)\citenamefont {Wu},
  \citenamefont {Chaunsali}, \citenamefont {Yasuda}, \citenamefont {Yu},\ and\
  \citenamefont {Yang}}]{wu2018dial}%
  \BibitemOpen
  \bibfield  {author} {\bibinfo {author} {\bibfnamefont {Y.}~\bibnamefont
  {Wu}}, \bibinfo {author} {\bibfnamefont {R.}~\bibnamefont {Chaunsali}},
  \bibinfo {author} {\bibfnamefont {H.}~\bibnamefont {Yasuda}}, \bibinfo
  {author} {\bibfnamefont {K.}~\bibnamefont {Yu}},\ and\ \bibinfo {author}
  {\bibfnamefont {J.}~\bibnamefont {Yang}},\ }\bibfield  {title} {\bibinfo
  {title} {Dial-in topological metamaterials based on bistable stewart
  platform},\ }\href@noop {} {\bibfield  {journal} {\bibinfo  {journal}
  {Scientific reports}\ }\textbf {\bibinfo {volume} {8}},\ \bibinfo {pages}
  {112} (\bibinfo {year} {2018})}\BibitemShut {NoStop}%
\bibitem [{\citenamefont {Pal}\ \emph {et~al.}(2016)\citenamefont {Pal},
  \citenamefont {Schaeffer},\ and\ \citenamefont {Ruzzene}}]{pal2016helical}%
  \BibitemOpen
  \bibfield  {author} {\bibinfo {author} {\bibfnamefont {R.~K.}\ \bibnamefont
  {Pal}}, \bibinfo {author} {\bibfnamefont {M.}~\bibnamefont {Schaeffer}},\
  and\ \bibinfo {author} {\bibfnamefont {M.}~\bibnamefont {Ruzzene}},\
  }\bibfield  {title} {\bibinfo {title} {Helical edge states and topological
  phase transitions in phononic systems using bi-layered lattices},\
  }\href@noop {} {\bibfield  {journal} {\bibinfo  {journal} {Journal of Applied
  Physics}\ }\textbf {\bibinfo {volume} {119}},\ \bibinfo {pages} {084305}
  (\bibinfo {year} {2016})}\BibitemShut {NoStop}%
\bibitem [{\citenamefont {Yasuda}\ \emph {et~al.}(2017)\citenamefont {Yasuda},
  \citenamefont {Tachi}, \citenamefont {Lee},\ and\ \citenamefont
  {Yang}}]{yasuda2017origami}%
  \BibitemOpen
  \bibfield  {author} {\bibinfo {author} {\bibfnamefont {H.}~\bibnamefont
  {Yasuda}}, \bibinfo {author} {\bibfnamefont {T.}~\bibnamefont {Tachi}},
  \bibinfo {author} {\bibfnamefont {M.}~\bibnamefont {Lee}},\ and\ \bibinfo
  {author} {\bibfnamefont {J.}~\bibnamefont {Yang}},\ }\bibfield  {title}
  {\bibinfo {title} {Origami-based tunable truss structures for non-volatile
  mechanical memory operation},\ }\href@noop {} {\bibfield  {journal} {\bibinfo
   {journal} {Nature communications}\ }\textbf {\bibinfo {volume} {8}},\
  \bibinfo {pages} {962} (\bibinfo {year} {2017})}\BibitemShut {NoStop}%
\bibitem [{\citenamefont {Slonczewski}\ and\ \citenamefont
  {Weiss}(1958)}]{slonczewski1958band}%
  \BibitemOpen
  \bibfield  {author} {\bibinfo {author} {\bibfnamefont {J.}~\bibnamefont
  {Slonczewski}}\ and\ \bibinfo {author} {\bibfnamefont {P.}~\bibnamefont
  {Weiss}},\ }\bibfield  {title} {\bibinfo {title} {Band structure of
  graphite},\ }\href@noop {} {\bibfield  {journal} {\bibinfo  {journal}
  {Physical review}\ }\textbf {\bibinfo {volume} {109}},\ \bibinfo {pages}
  {272} (\bibinfo {year} {1958})}\BibitemShut {NoStop}%
\bibitem [{\citenamefont {Chen}\ \emph {et~al.}(2019)\citenamefont {Chen},
  \citenamefont {Liu},\ and\ \citenamefont {Hu}}]{chen2019topological}%
  \BibitemOpen
  \bibfield  {author} {\bibinfo {author} {\bibfnamefont {Y.}~\bibnamefont
  {Chen}}, \bibinfo {author} {\bibfnamefont {X.}~\bibnamefont {Liu}},\ and\
  \bibinfo {author} {\bibfnamefont {G.}~\bibnamefont {Hu}},\ }\bibfield
  {title} {\bibinfo {title} {Topological phase transition in mechanical
  honeycomb lattice},\ }\href@noop {} {\bibfield  {journal} {\bibinfo
  {journal} {Journal of the Mechanics and Physics of Solids}\ }\textbf
  {\bibinfo {volume} {122}},\ \bibinfo {pages} {54} (\bibinfo {year}
  {2019})}\BibitemShut {NoStop}%
\bibitem [{\citenamefont {Mousavi}\ \emph {et~al.}(2015)\citenamefont
  {Mousavi}, \citenamefont {Khanikaev},\ and\ \citenamefont
  {Wang}}]{mousavi2015topologically}%
  \BibitemOpen
  \bibfield  {author} {\bibinfo {author} {\bibfnamefont {S.~H.}\ \bibnamefont
  {Mousavi}}, \bibinfo {author} {\bibfnamefont {A.~B.}\ \bibnamefont
  {Khanikaev}},\ and\ \bibinfo {author} {\bibfnamefont {Z.}~\bibnamefont
  {Wang}},\ }\bibfield  {title} {\bibinfo {title} {Topologically protected
  elastic waves in phononic metamaterials},\ }\href@noop {} {\bibfield
  {journal} {\bibinfo  {journal} {Nature communications}\ }\textbf {\bibinfo
  {volume} {6}},\ \bibinfo {pages} {8682} (\bibinfo {year} {2015})}\BibitemShut
  {NoStop}%
\bibitem [{\citenamefont {Mei}\ \emph {et~al.}(2012)\citenamefont {Mei},
  \citenamefont {Wu}, \citenamefont {Chan},\ and\ \citenamefont
  {Zhang}}]{mei2012first}%
  \BibitemOpen
  \bibfield  {author} {\bibinfo {author} {\bibfnamefont {J.}~\bibnamefont
  {Mei}}, \bibinfo {author} {\bibfnamefont {Y.}~\bibnamefont {Wu}}, \bibinfo
  {author} {\bibfnamefont {C.~T.}\ \bibnamefont {Chan}},\ and\ \bibinfo
  {author} {\bibfnamefont {Z.-Q.}\ \bibnamefont {Zhang}},\ }\bibfield  {title}
  {\bibinfo {title} {First-principles study of dirac and dirac-like cones in
  phononic and photonic crystals},\ }\href@noop {} {\bibfield  {journal}
  {\bibinfo  {journal} {Physical Review B}\ }\textbf {\bibinfo {volume} {86}},\
  \bibinfo {pages} {035141} (\bibinfo {year} {2012})}\BibitemShut {NoStop}%
\bibitem [{\citenamefont {Lu}\ \emph {et~al.}(2014{\natexlab{b}})\citenamefont
  {Lu}, \citenamefont {Qiu}, \citenamefont {Xu}, \citenamefont {Ye},
  \citenamefont {Ke},\ and\ \citenamefont {Liu}}]{lu2014dirac}%
  \BibitemOpen
  \bibfield  {author} {\bibinfo {author} {\bibfnamefont {J.}~\bibnamefont
  {Lu}}, \bibinfo {author} {\bibfnamefont {C.}~\bibnamefont {Qiu}}, \bibinfo
  {author} {\bibfnamefont {S.}~\bibnamefont {Xu}}, \bibinfo {author}
  {\bibfnamefont {Y.}~\bibnamefont {Ye}}, \bibinfo {author} {\bibfnamefont
  {M.}~\bibnamefont {Ke}},\ and\ \bibinfo {author} {\bibfnamefont
  {Z.}~\bibnamefont {Liu}},\ }\bibfield  {title} {\bibinfo {title} {Dirac cones
  in two-dimensional artificial crystals for classical waves},\ }\href@noop {}
  {\bibfield  {journal} {\bibinfo  {journal} {Physical Review B}\ }\textbf
  {\bibinfo {volume} {89}},\ \bibinfo {pages} {134302} (\bibinfo {year}
  {2014}{\natexlab{b}})}\BibitemShut {NoStop}%
\bibitem [{\citenamefont {Pikulin}\ \emph {et~al.}(2016)\citenamefont
  {Pikulin}, \citenamefont {Chen},\ and\ \citenamefont
  {Franz}}]{pikulin2016chiral}%
  \BibitemOpen
  \bibfield  {author} {\bibinfo {author} {\bibfnamefont {D.}~\bibnamefont
  {Pikulin}}, \bibinfo {author} {\bibfnamefont {A.}~\bibnamefont {Chen}},\ and\
  \bibinfo {author} {\bibfnamefont {M.}~\bibnamefont {Franz}},\ }\bibfield
  {title} {\bibinfo {title} {Chiral anomaly from strain-induced gauge fields in
  dirac and weyl semimetals},\ }\href@noop {} {\bibfield  {journal} {\bibinfo
  {journal} {Physical Review X}\ }\textbf {\bibinfo {volume} {6}},\ \bibinfo
  {pages} {041021} (\bibinfo {year} {2016})}\BibitemShut {NoStop}%
\bibitem [{\citenamefont {Grushin}\ \emph {et~al.}(2016)\citenamefont
  {Grushin}, \citenamefont {Venderbos}, \citenamefont {Vishwanath},\ and\
  \citenamefont {Ilan}}]{grushin2016inhomogeneous}%
  \BibitemOpen
  \bibfield  {author} {\bibinfo {author} {\bibfnamefont {A.~G.}\ \bibnamefont
  {Grushin}}, \bibinfo {author} {\bibfnamefont {J.~W.}\ \bibnamefont
  {Venderbos}}, \bibinfo {author} {\bibfnamefont {A.}~\bibnamefont
  {Vishwanath}},\ and\ \bibinfo {author} {\bibfnamefont {R.}~\bibnamefont
  {Ilan}},\ }\bibfield  {title} {\bibinfo {title} {Inhomogeneous weyl and dirac
  semimetals: Transport in axial magnetic fields and fermi arc surface states
  from pseudo-landau levels},\ }\href@noop {} {\bibfield  {journal} {\bibinfo
  {journal} {Physical Review X}\ }\textbf {\bibinfo {volume} {6}},\ \bibinfo
  {pages} {041046} (\bibinfo {year} {2016})}\BibitemShut {NoStop}%
\bibitem [{\citenamefont {Jia}\ \emph {et~al.}(2019)\citenamefont {Jia},
  \citenamefont {Zhang}, \citenamefont {Gao}, \citenamefont {Guo},
  \citenamefont {Yang}, \citenamefont {Hu}, \citenamefont {Bi}, \citenamefont
  {Xiang}, \citenamefont {Liu},\ and\ \citenamefont
  {Zhang}}]{jia2019observation}%
  \BibitemOpen
  \bibfield  {author} {\bibinfo {author} {\bibfnamefont {H.}~\bibnamefont
  {Jia}}, \bibinfo {author} {\bibfnamefont {R.}~\bibnamefont {Zhang}}, \bibinfo
  {author} {\bibfnamefont {W.}~\bibnamefont {Gao}}, \bibinfo {author}
  {\bibfnamefont {Q.}~\bibnamefont {Guo}}, \bibinfo {author} {\bibfnamefont
  {B.}~\bibnamefont {Yang}}, \bibinfo {author} {\bibfnamefont {J.}~\bibnamefont
  {Hu}}, \bibinfo {author} {\bibfnamefont {Y.}~\bibnamefont {Bi}}, \bibinfo
  {author} {\bibfnamefont {Y.}~\bibnamefont {Xiang}}, \bibinfo {author}
  {\bibfnamefont {C.}~\bibnamefont {Liu}},\ and\ \bibinfo {author}
  {\bibfnamefont {S.}~\bibnamefont {Zhang}},\ }\bibfield  {title} {\bibinfo
  {title} {Observation of chiral zero mode in inhomogeneous three-dimensional
  weyl metamaterials},\ }\href@noop {} {\bibfield  {journal} {\bibinfo
  {journal} {Science}\ }\textbf {\bibinfo {volume} {363}},\ \bibinfo {pages}
  {148} (\bibinfo {year} {2019})}\BibitemShut {NoStop}%
\bibitem [{\citenamefont {Jia}\ \emph {et~al.}(2022)\citenamefont {Jia},
  \citenamefont {Wang}, \citenamefont {Ma}, \citenamefont {Zhang},
  \citenamefont {Hu},\ and\ \citenamefont {Chan}}]{jia2022experimental}%
  \BibitemOpen
  \bibfield  {author} {\bibinfo {author} {\bibfnamefont {H.}~\bibnamefont
  {Jia}}, \bibinfo {author} {\bibfnamefont {M.}~\bibnamefont {Wang}}, \bibinfo
  {author} {\bibfnamefont {S.}~\bibnamefont {Ma}}, \bibinfo {author}
  {\bibfnamefont {R.-Y.}\ \bibnamefont {Zhang}}, \bibinfo {author}
  {\bibfnamefont {J.}~\bibnamefont {Hu}},\ and\ \bibinfo {author}
  {\bibfnamefont {C.~T.}\ \bibnamefont {Chan}},\ }\href@noop {} {\bibinfo
  {title} {Experimental realization of chiral landau levels in two-dimensional
  dirac cone systems with inhomogeneous effective mass}} (\bibinfo {year}
  {2022}),\ \Eprint {https://arxiv.org/abs/2209.10745} {arXiv:2209.10745
  [cond-mat.mtrl-sci]} \BibitemShut {NoStop}%
\bibitem [{\citenamefont {Taychatanapat}\ \emph {et~al.}(2015)\citenamefont
  {Taychatanapat}, \citenamefont {Tan}, \citenamefont {Yeo}, \citenamefont
  {Watanabe}, \citenamefont {Taniguchi},\ and\ \citenamefont
  {{\"O}zyilmaz}}]{taychatanapat2015conductance}%
  \BibitemOpen
  \bibfield  {author} {\bibinfo {author} {\bibfnamefont {T.}~\bibnamefont
  {Taychatanapat}}, \bibinfo {author} {\bibfnamefont {J.~Y.}\ \bibnamefont
  {Tan}}, \bibinfo {author} {\bibfnamefont {Y.}~\bibnamefont {Yeo}}, \bibinfo
  {author} {\bibfnamefont {K.}~\bibnamefont {Watanabe}}, \bibinfo {author}
  {\bibfnamefont {T.}~\bibnamefont {Taniguchi}},\ and\ \bibinfo {author}
  {\bibfnamefont {B.}~\bibnamefont {{\"O}zyilmaz}},\ }\bibfield  {title}
  {\bibinfo {title} {Conductance oscillations induced by ballistic snake states
  in a graphene heterojunction},\ }\href@noop {} {\bibfield  {journal}
  {\bibinfo  {journal} {Nature communications}\ }\textbf {\bibinfo {volume}
  {6}},\ \bibinfo {pages} {6093} (\bibinfo {year} {2015})}\BibitemShut
  {NoStop}%
\bibitem [{\citenamefont {Oroszl{\'a}ny}\ \emph {et~al.}(2008)\citenamefont
  {Oroszl{\'a}ny}, \citenamefont {Rakyta}, \citenamefont {Korm{\'a}nyos},
  \citenamefont {Lambert},\ and\ \citenamefont {Cserti}}]{oroszlany2008theory}%
  \BibitemOpen
  \bibfield  {author} {\bibinfo {author} {\bibfnamefont {L.}~\bibnamefont
  {Oroszl{\'a}ny}}, \bibinfo {author} {\bibfnamefont {P.}~\bibnamefont
  {Rakyta}}, \bibinfo {author} {\bibfnamefont {A.}~\bibnamefont
  {Korm{\'a}nyos}}, \bibinfo {author} {\bibfnamefont {C.}~\bibnamefont
  {Lambert}},\ and\ \bibinfo {author} {\bibfnamefont {J.}~\bibnamefont
  {Cserti}},\ }\bibfield  {title} {\bibinfo {title} {Theory of snake states in
  graphene},\ }\href@noop {} {\bibfield  {journal} {\bibinfo  {journal}
  {Physical Review B}\ }\textbf {\bibinfo {volume} {77}},\ \bibinfo {pages}
  {081403} (\bibinfo {year} {2008})}\BibitemShut {NoStop}%
\bibitem [{\citenamefont {Ghosh}\ \emph {et~al.}(2008)\citenamefont {Ghosh},
  \citenamefont {De~Martino}, \citenamefont {H{\"a}usler}, \citenamefont
  {Dell’Anna},\ and\ \citenamefont {Egger}}]{ghosh2008conductance}%
  \BibitemOpen
  \bibfield  {author} {\bibinfo {author} {\bibfnamefont {T.}~\bibnamefont
  {Ghosh}}, \bibinfo {author} {\bibfnamefont {A.}~\bibnamefont {De~Martino}},
  \bibinfo {author} {\bibfnamefont {W.}~\bibnamefont {H{\"a}usler}}, \bibinfo
  {author} {\bibfnamefont {L.}~\bibnamefont {Dell’Anna}},\ and\ \bibinfo
  {author} {\bibfnamefont {R.}~\bibnamefont {Egger}},\ }\bibfield  {title}
  {\bibinfo {title} {Conductance quantization and snake states in graphene
  magnetic waveguides},\ }\href@noop {} {\bibfield  {journal} {\bibinfo
  {journal} {Physical Review B}\ }\textbf {\bibinfo {volume} {77}},\ \bibinfo
  {pages} {081404} (\bibinfo {year} {2008})}\BibitemShut {NoStop}%
\bibitem [{\citenamefont {M{\"u}ller}(1992)}]{muller1992effect}%
  \BibitemOpen
  \bibfield  {author} {\bibinfo {author} {\bibfnamefont {J.}~\bibnamefont
  {M{\"u}ller}},\ }\bibfield  {title} {\bibinfo {title} {Effect of a nonuniform
  magnetic field on a two-dimensional electron gas in the ballistic regime},\
  }\href@noop {} {\bibfield  {journal} {\bibinfo  {journal} {Physical review
  letters}\ }\textbf {\bibinfo {volume} {68}},\ \bibinfo {pages} {385}
  (\bibinfo {year} {1992})}\BibitemShut {NoStop}%
\end{thebibliography}%

\end{document}